\newif\ifIEEEtranAvailable
\IEEEtranAvailabletrue      

\ifIEEEtranAvailable
  \documentclass[journal]{IEEEtran}
\else
  \documentclass[10pt,twocolumn]{article}
  \usepackage[margin=0.75in]{geometry}
  \newcommand{\IEEEPARstart}[2]{\textbf{\Large #1}\textsc{#2}}
  
\fi

\usepackage{amsmath,amssymb}
\usepackage{graphicx}
\usepackage{booktabs}
\usepackage{url}
\usepackage{xcolor}
\usepackage{tabularx}
\usepackage{array}
\usepackage{verbatim}
\usepackage{graphicx}
\usepackage{cite}
\usepackage{lipsum}
\usepackage{subcaption}
\usepackage[numbers,sort&compress]{natbib}
\newcommand{\figstub}[4]{%
  \IfFileExists{#1}{\includegraphics[width=#4\columnwidth]{#1}}{%
    \fbox{\begin{minipage}[c][0.5\columnwidth][c]{#4\columnwidth}
      \centering
      \small\ttfamily PLACEHOLDER

      \detokenize{#1}

      \rmfamily\itshape\footnotesize
      (place this file alongside main.tex to replace)
    \end{minipage}}%
  }%
}

\begin{document}

\title{Short-Range Microwave Interprobe Coupling in a Brain-Mimicking Phantom and Ex-Vivo Brain}
\author{Pouya Faeghi, Xiaoting Jia, and Jeffrey S. Walling%
\thanks{This research received no external funding.
Pouya Faeghi, Xiaoting Jia, and Jeffrey S. Walling are with the Bradley
Department of Electrical and Computer Engineering, Virginia Tech,
Blacksburg, VA, USA (e-mail: pfaeghi, xjia, jswalling@vt.edu).}}
\markboth{IEEE Journal of Electromagnetics, RF and Microwaves in Medicine and Biology}%
{Faeghi \MakeLowercase{\textit{et al.}}: Short-Range Microwave Interprobe Coupling in Brain Phantom and Ex Vivo Brain}
\newcommand{\absub}[1]{\textbf{#1:}~}
\IEEEaftertitletext{%
\vspace{-0.4\baselineskip}
{\small\noindent\textbf{\textit{Abstract}}---%
\absub{Objectives}Short-range wireless links (microwave range) between closely spaced devices in brain tissue are relevant to neural interfaces and intracranial sensing, yet the change in
coupling between two such devices caused by a finite volume of brain-like tissue has rarely been measured directly. \absub{Technology or Method}This work characterizes a printed,
microstrip-fed, parasitic-strip-loaded tapered-slot coupling probe in which only a partially inserted 12-mm wire enters the medium, while the printed structure remains outside. \absub{
Results}Two fabricated probes showed a measured free-space $-10$-dB reflection band of 5.78 to 6.86~GHz and a fractional bandwidth of approximately 17.1\%, consistent with full-wave
simulation. In a controlled two-probe experiment, a homogeneous agar--gelatin--saline brain-mimicking phantom was introduced and removed while the probe arrangement and measurement
conditions were retained. Introducing the phantom increased the measured transmission coefficient near 6.3~GHz by approximately 16~dB, and this differential was reproduced across repeated
trials. More than five measurements in ex vivo goat brain followed the phantom transmission response within approximately 1.5~dB across the 5.78--6.86-GHz band. Time-domain analysis of
the phantom-present link gave a system root-mean-square delay spread of approximately 210~ps. Full-wave dosimetry predicted a peak 1-g-averaged specific absorption rate of 1.05~W/kg for
approximately 0.49~W accepted power. \absub{Conclusions and Clinical or Biological Impact}These results provide an experimentally validated reference for short-range probe coupling in
brain-like media and support phantom-based bench evaluation of minimally invasive intracranial probes before in vivo studies.\par}
\vspace{0.5\baselineskip}
{\small\bfseries\noindent\textit{Keywords}---Biological tissues,
biomedical telemetry, brain phantoms, ex vivo measurements, microwave
coupling, near-field coupling, probe antennas, tapered-slot antennas.\par}
\vspace{0.8\baselineskip}
}
\maketitle

\section{Introduction}
\label{sec:intro}

\IEEEPARstart{S}{hort-range} wireless links are being investigated for
implantable and neural interfaces that require increased data throughput
within constrained form factors. Reviews of regular antennas and implantable antennas highlight
the competing requirements imposed by miniaturization, bandwidth,
biocompatibility, and the surrounding biological medium
\cite{implantreview2024, 11104879, d52}. High-data-rate impulse-radio transmitters for
brain interfaces further illustrate the demand for broadband telemetry
\cite{song2022,ding2024}.

Biological tissues are dispersive and lossy, with relative permittivity
and conductivity varying with frequency and tissue type
\cite{gabriel1996ii, 10732534}. Permittivity and conductivity influence propagation and loading, but they do not alone determine the measured response of a finite probe–medium–probe configuration. 
In a two-probe configuration, measured $S_{21}$ depends on the complete probe--medium electromagnetic system, including loading, mutual coupling, interfaces, geometry, propagation, and material loss. Net transmission therefore cannot be inferred from bulk tissue attenuation alone.

Previous in-body studies have measured or modeled path loss in
homogeneous tissues and living-animal configurations
\cite{kurup2012,chavez2015,floor2015}. Brain-implantable antennas have
also been characterized in multilayer head phantoms
\cite{hout2019,sapari2022}. Kamel \emph{et al.} measured
implant-to-on-body $S_{21}$ at 5.81~GHz with an implantable transmitter
immersed in chicken tissue and an on-body receiver placed on the tissue
surface; a prepared solution was evaluated separately for
implantable-antenna $S_{11}$ characterization~\cite{dualband2025}.
Related intracranial-pressure telemetry work has characterized
implantable antennas at 2.45~GHz using reflection and transmission
measurements through a scalp phantom~\cite{warty2008}. Neither study
performed the phantom-present/phantom-absent comparison used here. 

Existing studies have primarily characterized in-body path loss, implanted antennas, or transmission through homogeneous tissue models. They do not directly quantify the change in short-range two-probe coupling produced by introducing a finite, brain-mimicking medium while retaining the probe geometry.
The comparison controls the principal measurement-analysis variables and geometry, but it does not uniquely separate loading, mismatch, mutual coupling, propagation, interfaces, finite boundaries, and material loss. The result is therefore interpreted at the complete-system level rather than as a measurement of an intrinsic material mechanism.

The contributions are a microstrip-fed, parasitic-strip-loaded
tapered-slot coupling probe with a partially inserted 12-mm wire;
simulated and measured free-space $S_{11}$ characterization;
complementary numerical evidence from a single-probe human-head/brain
model; a controlled phantom-present/phantom-absent $S_{21}$ comparison
showing an approximately $+16$~dB differential near 6.3~GHz;
more than five repeated ex vivo goat-brain measurements; and time-domain
characterization of the phantom-present link. 

\section{Methods and Procedures}
\subsection{Probe Architecture and Design}
\label{sec:design}

\subsubsection{Architecture}

The device is a microstrip-fed, parasitic-strip-loaded tapered-slot
coupling probe with a partially inserted wire element. The printed
structure remains outside the investigated medium, while the wire is the
only conductive element inserted into it. This arrangement permits the
probe--medium interaction to be studied without immersing the full printed
board. The shorter term ``printed tapered-slot coupling probe'' is used
below.

\begin{figure}[t]
\centering
\includegraphics[width=8.9cm]{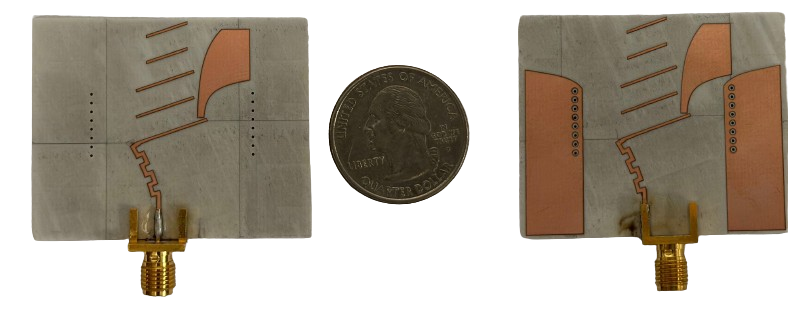}
\caption{Fabricated printed tapered-slot coupling probe. The left figure shows the bottom layer, and the right shows the top layer.}
\label{fig:probe_fabrication}
\end{figure}

The probe was fabricated on a $45 \times 37$~mm$^2$ RO4003C laminate with a
thickness of 0.508~mm, relative permittivity of 3.55, and loss tangent of
0.0027. Copper metallization is present on both faces. An edge-mounted  SubMiniature version A (SMA)
connector excites a 1.1-mm-wide microstrip line followed by stepped-width
sections and a tapered slot. Three rectangular parasitic strips are
located near the wider end of the taper. These descriptors identify the
geometry;
Fig.~\ref{fig:probe_fabrication} shows the fabricated
probe and its principal geometric features.
\subsubsection{Partially Inserted Wire Element}

A copper wire of 0.50~mm diameter and 12~mm exposed length is soldered to
one parasitic strip and extends beyond the board edge. The wire is not
connected directly to the microstrip feed; it is coupled through the
printed structure. During the loaded measurements, approximately 9~mm of
the wire was inserted into the investigated medium and approximately
3~mm, together with the complete printed structure, remained outside.

The 12-mm length is close to one quarter of the free-space wavelength at
the approximately 6.32-GHz band center.
\subsubsection{Dimensional Summary}

Table~\ref{tab:dims} summarizes the documented probe dimensions. The
dimensions describe the fabricated prototype and do not assign a unique
mechanism to any individual feature.

\begin{table}[t]
\caption{Documented dimensions of the coupling probe}
\label{tab:dims}
\centering
\begin{tabular}{lll}
\toprule
Parameter & Description & Value (mm) \\
\midrule
$L_s$ & Substrate length & 45.0 \\
$W_s$ & Substrate width & 37.0 \\
$h$   & Substrate thickness & 0.508 \\
$W_f$ & Feed width & 1.1 \\
$L_t$ & Taper length & 22.5 \\
$W_a$ & Taper opening & 18.0 \\
$L_d$ & Parasitic-strip length & 12.5 \\
$W_d$ & Parasitic-strip width & 8.0 \\
$g_d$ & Inter-strip gap & 1.5 \\
$L_w$ & Exposed wire length & 12.0 \\
$d_w$ & Wire diameter & 0.50 \\
\bottomrule
\end{tabular}
\end{table}

\subsection{Free-Space Characterization and Numerical Modeling}
\label{sec:freespace}

\subsubsection{Free-Space Reflection Response}

\begin{figure}[htbp]
\centering
\includegraphics[width=0.5\textwidth]{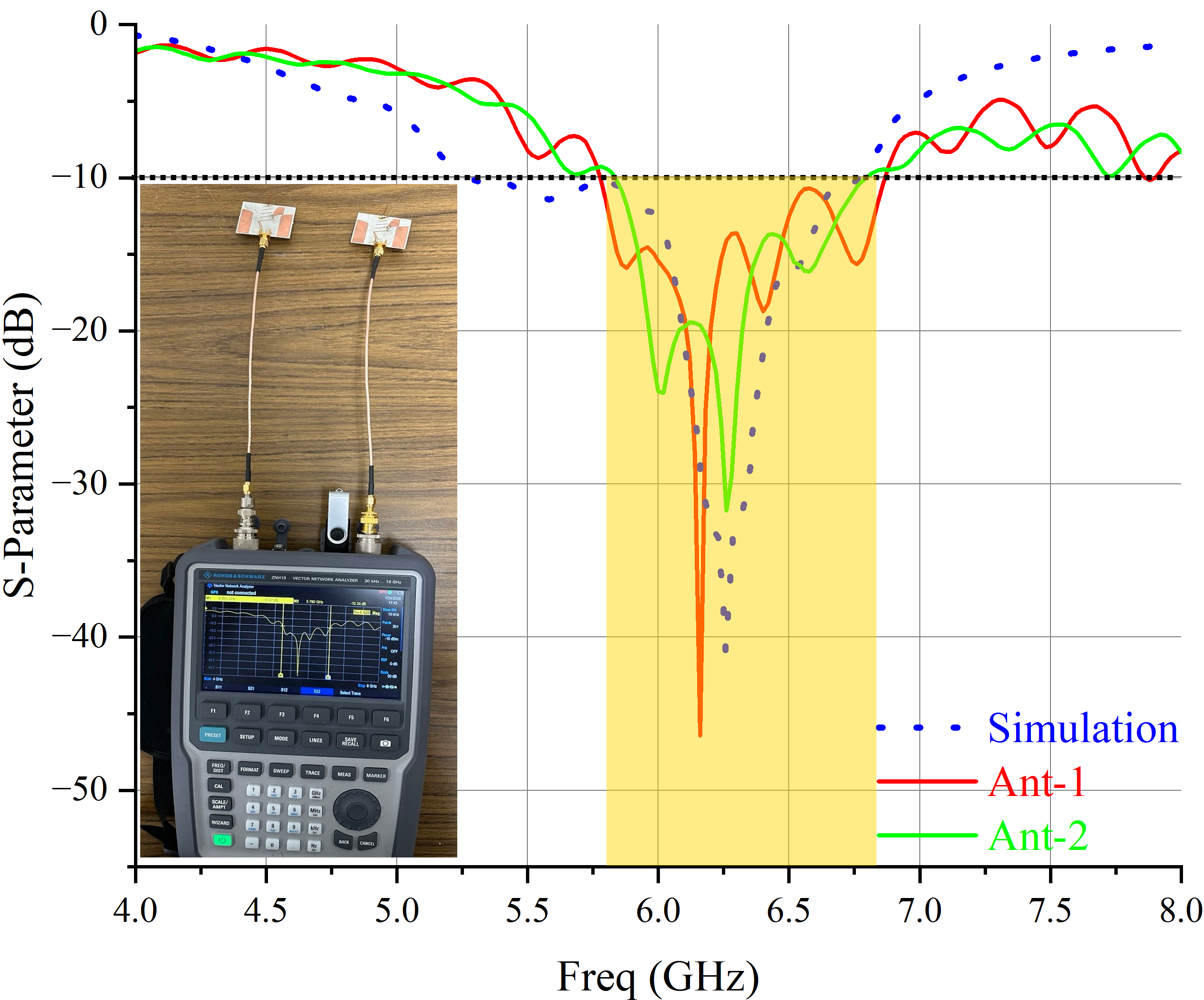}
\caption{Simulated and measured free-space reflection coefficients for
two fabricated probes. The measured $-10$ dB band extends from 5.78 to 6.86~GHz. Left: Ant-1, Right: Ant- 2}
\label{fig:s11_freespace}
\end{figure}

The free-space reflection response was measured for two independently
fabricated probes using the instrumentation and reference-plane procedure
described in Section~\ref{sec:setup}. Fig.~\ref{fig:s11_freespace} compares
the measured responses with the full-wave simulation. Both measured units
exhibit a $-10$-dB impedance band from 5.78 to 6.86~GHz. The corresponding
bandwidth is 1.08~GHz, the center frequency is approximately 6.32~GHz,
and the fractional bandwidth is approximately 17.1$\%$.

Ant~1 reaches a minimum of approximately $-46$~dB at 6.16~GHz, and
Ant~2 reaches approximately $-32$~dB at 6.27~GHz. The simulated minimum
is approximately $-41$~dB at 6.28~GHz. The agreement in band edges and
the proximity of the measured and simulated minima support use of the
model for complementary numerical characterization. 

\subsubsection{Simulated Radiation Characteristics}
\begin{figure}[t]
    \centering
    \begin{tabular}{cc}
        \begin{subfigure}{0.45\linewidth}
            \centering
            \includegraphics[width=\linewidth]{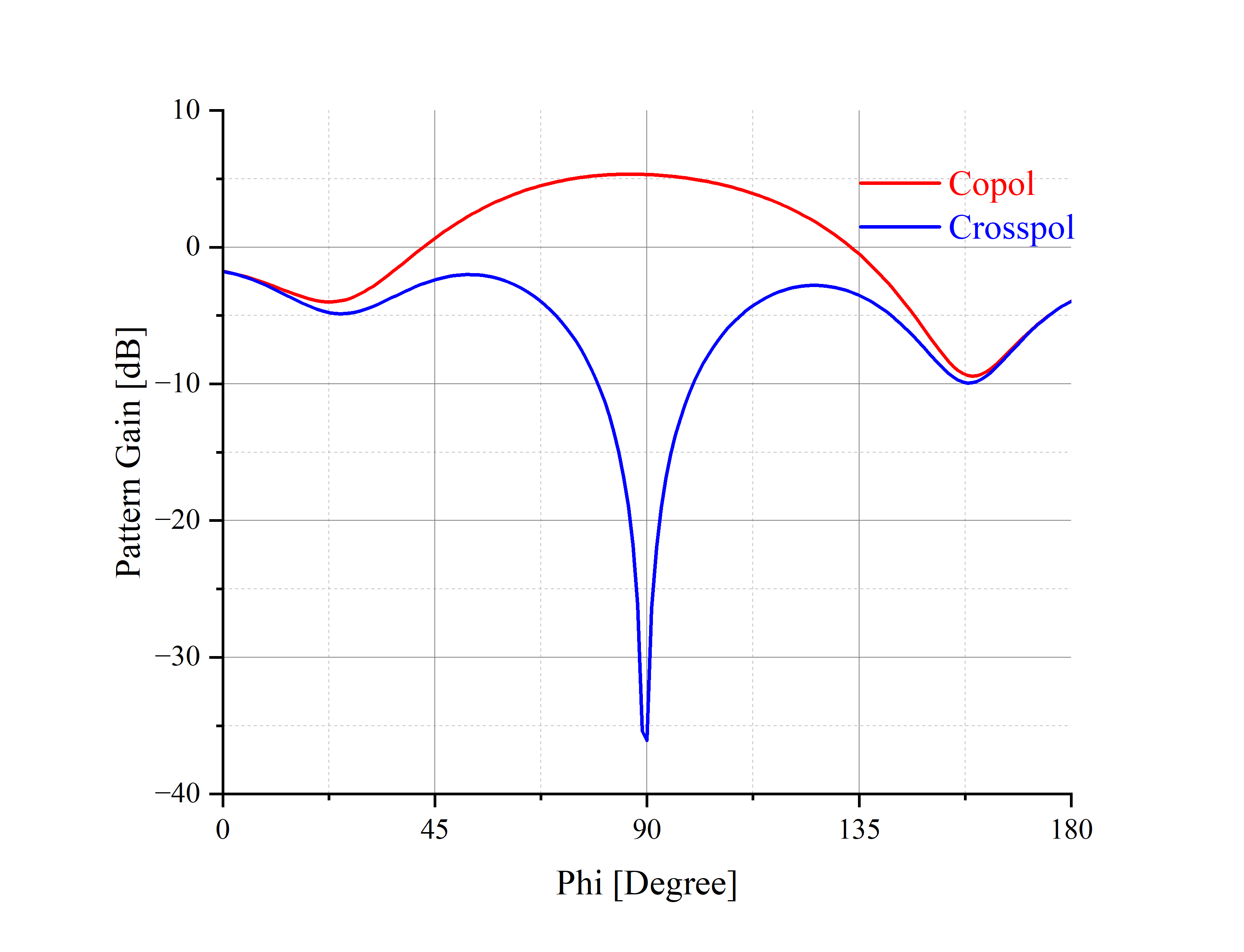}
            \caption{Phi Cartesian}
            \label{subfig:phi_cartesian}
        \end{subfigure}
        &
        \begin{subfigure}{0.45\linewidth}
            \centering
            \includegraphics[width=\linewidth]{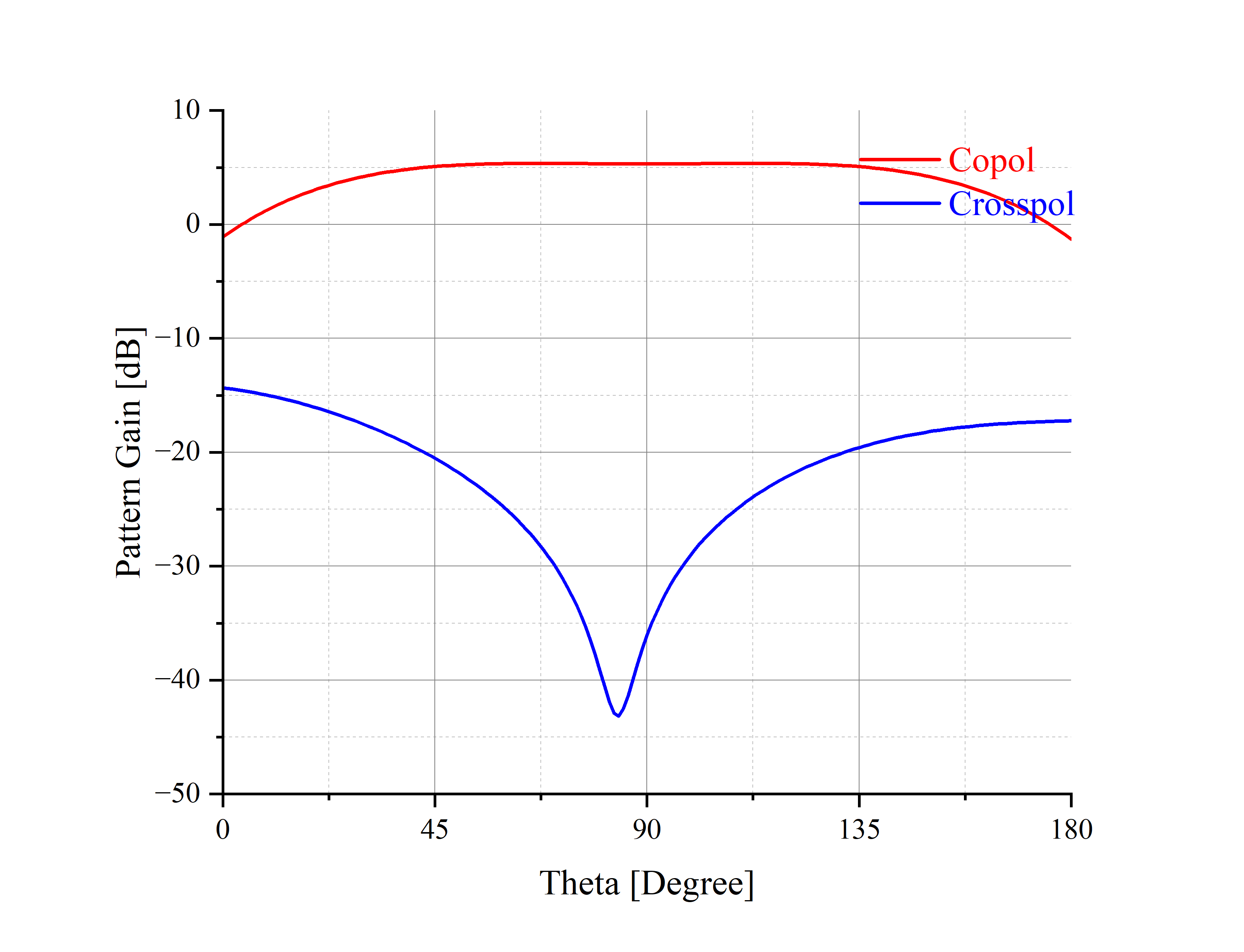}
            \caption{Theta Cartesian}
            \label{subfig:theta_cartesian}
        \end{subfigure}
    \end{tabular}
    \vspace{0.3cm}
    \begin{subfigure}{0.9\linewidth}
        \centering
        \includegraphics[width=\linewidth]{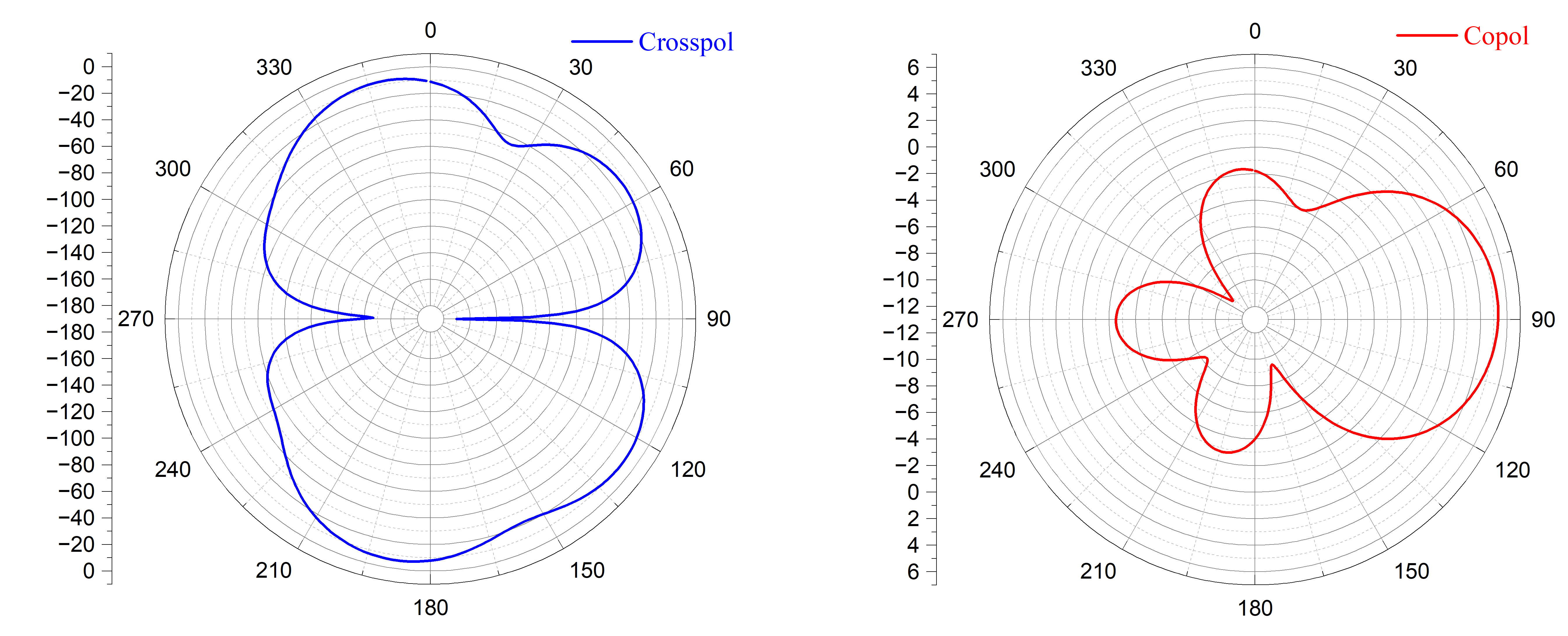}
        \caption{Phi Polar}
        \label{subfig:phi_polar}
    \end{subfigure}
    \vspace{0.3cm}
    \begin{subfigure}{0.9\linewidth}
        \centering
        \includegraphics[width=\linewidth]{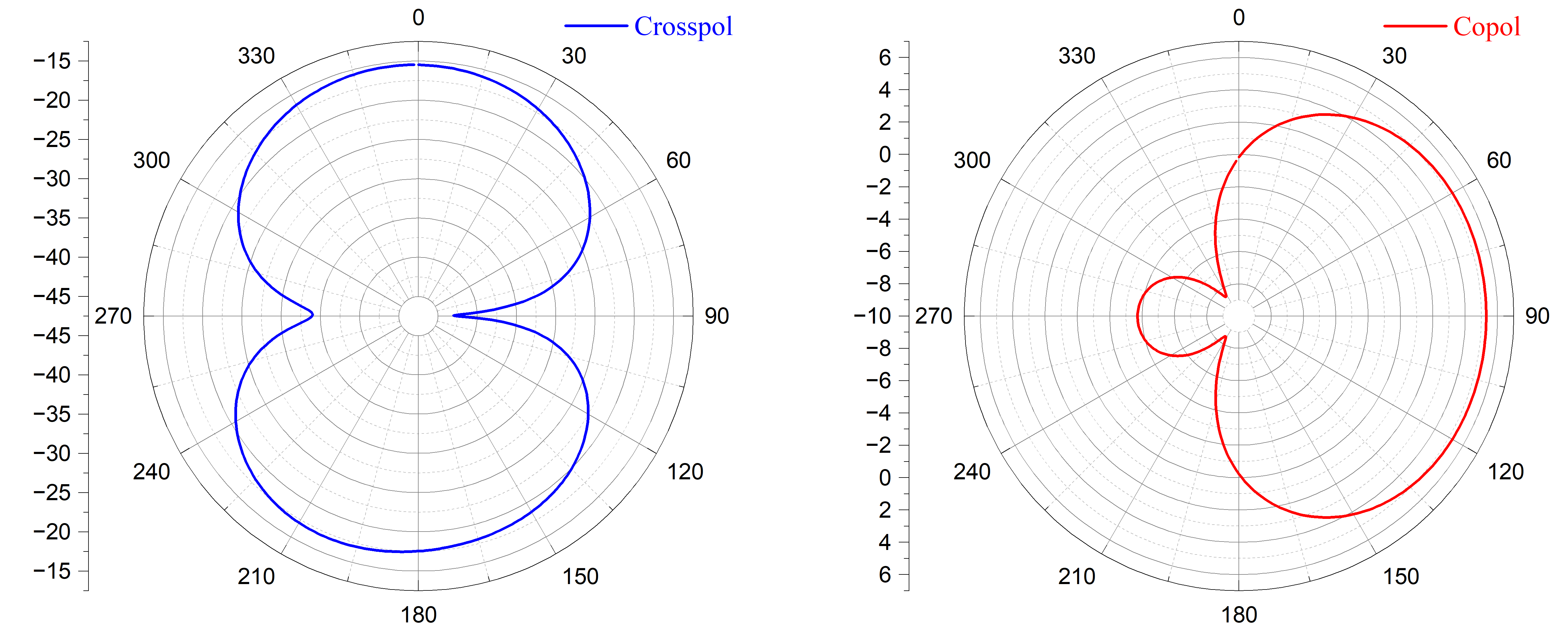}
        \caption{Theta Polar}
        \label{subfig:theta_polar}
    \end{subfigure}
    \caption{Radiation patterns at 6.3 GHz showing the co-polarized and cross-polarized components in the principal planes.}
    \label{fig:patterns}
\end{figure}

The radiation quantities in this subsection are simulation results.
The simulated free-space characteristics at 6.3~GHz are summarized in
Table~\ref{tab:rad}, and the simulated principal-plane patterns are shown
in Fig.~\ref{fig:patterns}. These free-space quantities characterize the
numerical model.

\begin{table}[t]
\caption{Simulated free-space radiation characteristics at 6.3~GHz}
\label{tab:rad}
\centering
\begin{tabular}{ll}
\toprule
Quantity & Simulated value \\
\midrule
Realized gain & 5.34 dBi \\
Radiation efficiency & 98\% \\
Half-power beamwidth & 64$^\circ$ \\
Front-to-back ratio & 9.8 dB \\
\bottomrule
\end{tabular}
\end{table}
\subsubsection{Two-Probe Numerical Support Model}

\begin{figure}[htbp]
\centering
\includegraphics[width=0.5\textwidth]{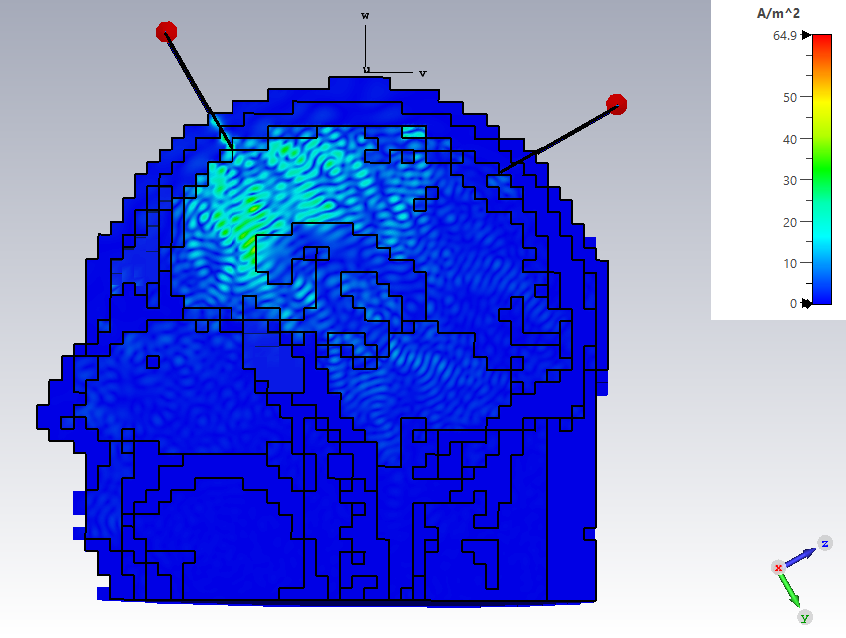}
\caption{Current-density distribution at 6.3 GHz in the two-probe human-head/brain simulation}
\label{fig:surface_current}
\end{figure}

An earlier CST Studio Suite human-head/brain model containing two printed probes and
their partially inserted wire was used as complementary numerical support
for probe--tissue interaction. Fig.~\ref{fig:surface_current} shows the
simulated current-density distribution at 6.3~GHz. The distribution demonstrates a spatially nonuniform current density within the head/brain model in the vicinity of the probe–tissue interaction region and throughout portions of the surrounding tissue. The image is interpreted qualitatively. Therefore, this result is used to support the presence and spatial distribution of electromagnetic interaction between the probes and the tissue.

The same two-probe transmit–receive configuration was subsequently used for the specific absorption rate (SAR) assessment presented in 
Section~\ref{sec:safety}.
\subsection{Experimental Media and Measurement Configuration}
\label{sec:setup}

\subsubsection{Brain-Mimicking Agar--Gelatin--Saline Phantom}
The experimental medium was a homogeneous, brain-mimicking
agar--gelatin--saline phantom made with deionized water. The formulation was
adapted from the simplified gelatin--saline compositions examined in
Series~1 and Series~2~\cite{d1}. Agar was additionally introduced as a
structural gelling agent to improve mechanical stability and shape retention.
Gelatin and agar were dissolved in deionized water under controlled heating
and continuous magnetic stirring until a homogeneous mixture was obtained,
while saline was added to adjust the electrical conductivity. The
prepared solution was subsequently poured into the desired mold and allowed
to cool and solidify. This simplified formulation was selected to provide a
reproducible and mechanically stable medium for electromagnetic experiments.

Since the phantom is primarily water, its microwave response depends on dipolar relaxation of water molecules and ionic conduction due to dissolved salts. Gelatin--NaCl--water gels have been characterized from 0.2
to 20~GHz and follow a Cole--Cole dispersion whose permittivity and loss are set by the gelatin and salt concentrations~\cite{buff2006}, placing them in
the range of high-water-content tissues such as brain gray matter
($\varepsilon_r' = 41.0$, $\sigma = 4.65$~S/m at 6~GHz)~\cite{nict}.



\subsubsection{Two-Probe Phantom and Air-Control Configuration}
A Rohde \& Schwarz ZNH18 vector network analyzer (VNA) was used to conduct measurements across 4--8 GHz, utilizing 201 points, a 20-MHz frequency increment, a 10-kHz intermediate-frequency bandwidth, and a source power of  $-10$~dBm.

A
short--open--load--through calibration was performed with a Rohde \&
Schwarz ZN-Z170 mechanical calibration kit at the VNA instrument ports
before the radio-frequency (RF) cables were attached. The cables therefore lie outside the
calibrated reference planes. No cable de-embedding or port extension was
applied. The same cables, instrument settings, and arrangement were
retained between the controlled phantom-present and phantom-absent
measurements. For each comparison, the same support arrangement, cable
configuration, and probe placement were deliberately retained.

Two printed probes faced one another with a nominal relative angle of
$45^{\circ} \pm 10^{\circ}$. The
tip-to-tip spacing was approximately 9--12~mm. 
Approximately 9~mm of each 12-mm wire was inserted
into the phantom, while approximately 3~mm and the complete printed
structures remained outside the medium.

For the phantom-present condition, the phantom bridged the inserted wire
regions. For the air control, the phantom was removed while the 
probe arrangement and measurement conditions were retained. This control
measures the change in the complete two-probe response associated with
introducing the finite medium.
The overall arrangement is shown
in Fig. \ref{fig:phantom_air_control}, \ref{fig:goat_setup}.

The phantom-present and air-control comparison was repeated with the same arrangement, and each trial showed similar $S_{21}$ differential behavior.
\begin{figure}[htbp]
\centering
\includegraphics[width=0.5\textwidth]{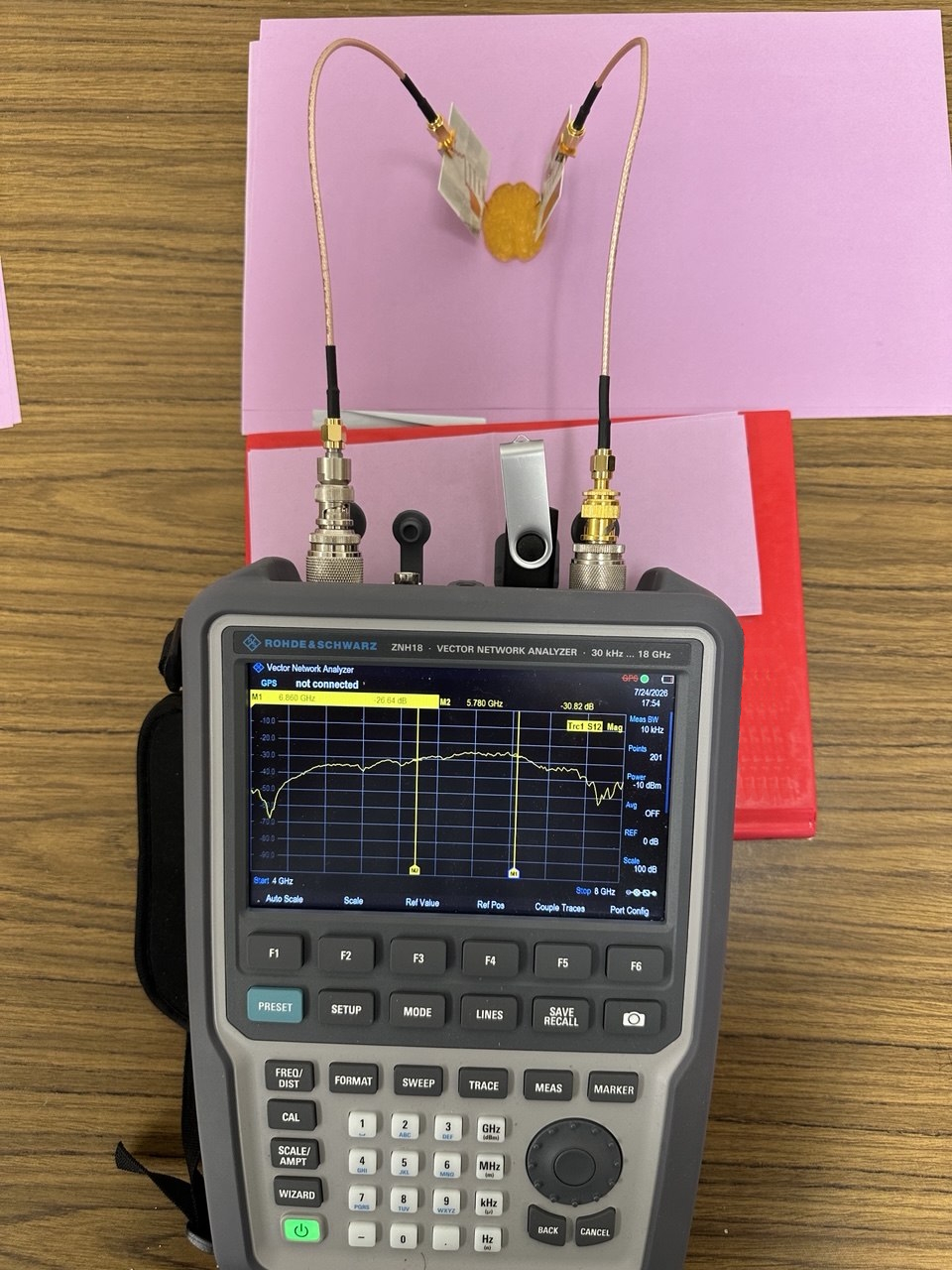}
\caption{Two-probe measurement configuration $S_{21}$ with the phantom bridging the partially inserted wires.}
\label{fig:phantom_loaded}
\end{figure}

\subsubsection{\textit{Ex vivo} Goat-Brain Validation}

The coupling measurement was repeated with ex vivo goat brain using the
same ZNH18 VNA, probes, measurement cables, support, instrument settings,
and nominal geometry as closely as practicable. A minimum of five repeated
measurements were performed. The repeated measurements exhibited the
same general $S_{11}$ and $S_{21}$ behavior. These measurements provide
repeated ex vivo validation of the operating-band response rather than
formal statistical analysis. They do not establish equivalence to
human or in vivo tissue or identity of the underlying coupling mechanisms.
Fig.~\ref{fig:goat_setup} documents the ex vivo measurement arrangement.

\begin{figure}[t]
\centering
\figstub{goat_brain.jpg}{}{}{1.0}
\caption{\textit{Ex vivo} goat-brain measurement arrangement using the same VNA,
probes, cables, support, settings, and nominal geometry as the phantom
measurements as closely as practicable.}
\label{fig:goat_setup}
\end{figure}

\subsubsection{Relationship Between Numerical and Experimental Configurations}

Two distinct numerical configurations support this work. The
single-probe CST human-head/brain model described in
Section~\ref{sec:freespace} provides complementary information about
probe--tissue interaction and surface-current distribution.
A separate two-probe CST configuration was used for the numerical SAR
assessment in Section~\ref{sec:safety}. In that model, one probe was
excited as the transmitter and the opposite probe was matched as the
receiver. 
\subsubsection{Scope and Limitations}
\label{subsec:scope}
The experiment represents a finite, partially inserted two-probe
configuration rather than a fully implanted link. The printed structures
remain in air; the medium has finite boundaries; the phantom is
homogeneous.

\section{Results}  
\subsection{Inter-Probe Coupling}
\label{sec:coupling}

\subsubsection{Phantom-Present Response}

With the phantom bridging the partially inserted wires, the measured
$|S_{21}|$ forms a broad plateau over the principal operating band. Near
6.3~GHz, $|S_{21}|$ is approximately $-24$~dB. The
phantom-present $|S_{11}|$ remains below $-10$~dB near this frequency.
The phantom-loaded geometry is documented in
Fig.~\ref{fig:phantom_loaded}.

\subsubsection{Phantom-Absent Control}

\begin{figure}[htbp]
\centering
\includegraphics[width=0.5\textwidth]{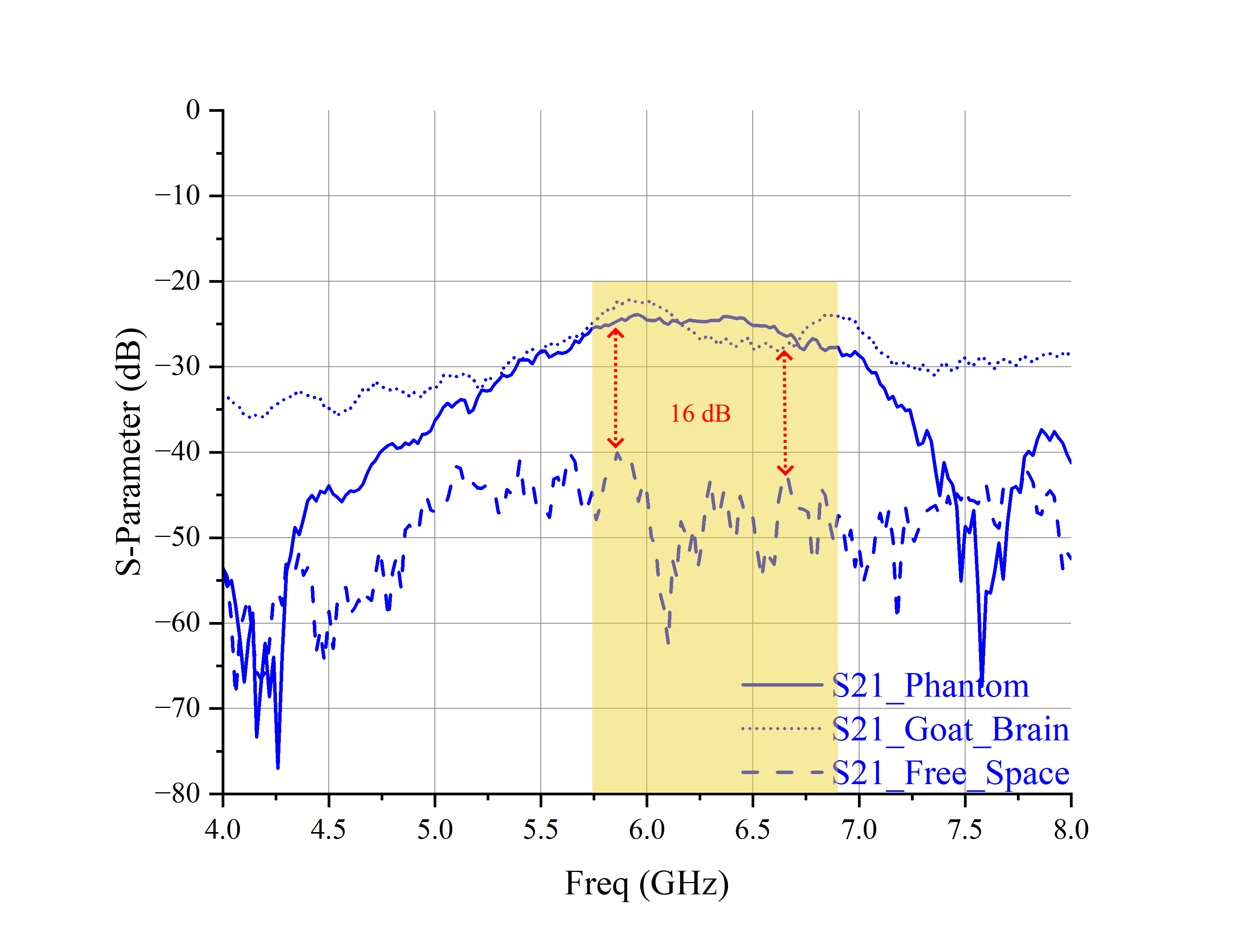}
\caption{Measured transmission with the phantom present and absent while
retaining the nominal geometry and measurement conditions. The differential
near 6.3~GHz is approximately $+16$~dB.}
\label{fig:phantom_air_control}
\end{figure}

Removing the phantom while retaining the probe arrangement and
measurement analysis reduces $|S_{21}|$ near 6.3~GHz to approximately
$-40$~dB. The phantom-present minus phantom-absent difference is therefore
approximately $+16$~dB in this frequency region, as shown in
Fig.~\ref{fig:phantom_air_control}. Defining
\begin{equation}
\Delta S_{21,\mathrm{dB}}
=
S_{21,\mathrm{phantom,dB}}-S_{21,\mathrm{air,dB}},
\label{eq:delta}
\end{equation}
a 16-dB difference in $S_{21}$ corresponds to approximately a 39.8-fold difference in $|S_{21}|^2$. Because the measurements are based on VNA scattering parameters rather than calibrated delivered power, this ratio should not be interpreted as an absolute received-power or link-budget ratio.

The phantom-present/air-control comparison was repeated under the same
nominal retained arrangement, with the same general $S_{21}$ differential
behavior observed across repetitions. The reported  $+16$~dB
differential remains the representative principal value.

\subsubsection{\textit{Ex vivo} Goat-Brain Response}

\begin{figure}[htbp]
\centering
\includegraphics[width=0.5\textwidth]{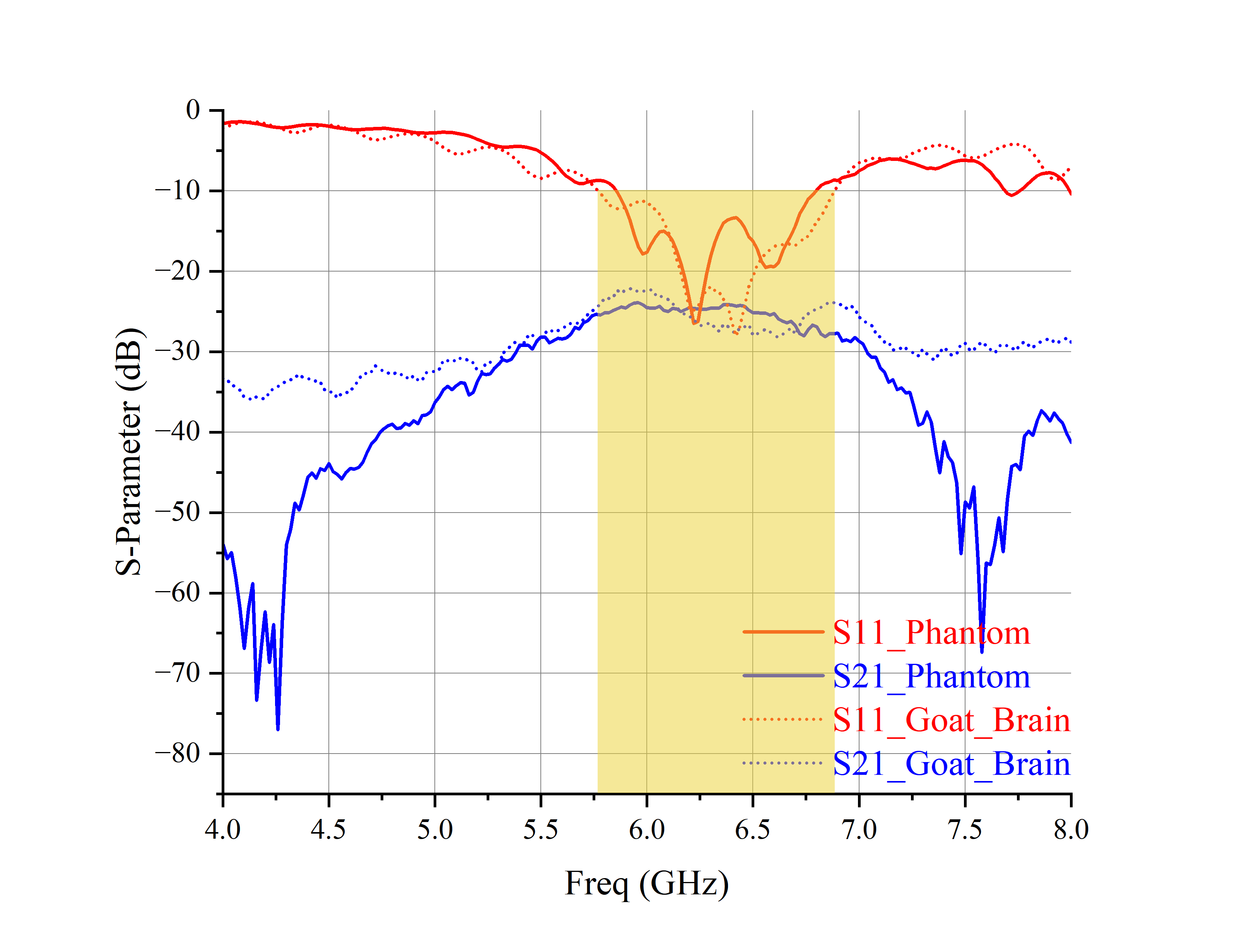}
\caption{Representative phantom and ex vivo goat-brain $S_{11}$ and
$S_{21}$ responses.} 
\label{fig:goat_comparison}
\end{figure}

Fig.~\ref{fig:goat_comparison} compares the phantom and ex vivo goat-brain
measurements. Across the 5.78--6.86-GHz operating band, the goat-brain
$S_{21}$ response closely follows the phantom response. The largest
visually apparent difference within this band is approximately 1.5~dB.
The responses diverge more strongly above
approximately 7~GHz.

More than five goat-brain measurements were performed using the same VNA,
probes, cables, support, settings, and nominal geometry as closely as
practicable, with reproducible general $S_{11}$ and $S_{21}$ behavior.
The repeated measurements provide ex vivo validation of the
operating-band response.

\subsubsection{Time-Domain Channel Response}

The phantom-present complex $S_{21}$ record spans 4--8~GHz with
$N=201$ samples and $\Delta f=20$~MHz. A Hann window was applied before a
direct inverse discrete Fourier transform with eightfold zero padding.
The 4-GHz measured span gives a nominal Fourier delay resolution of
$1/B=250$~ps, while the frequency spacing gives a 50-ns delay interval.
Delay statistics were evaluated over a 0--10-ns gate using a $-20$-dB
power threshold.

\begin{figure}[htbp]
\centering
\includegraphics[width=0.5\textwidth]{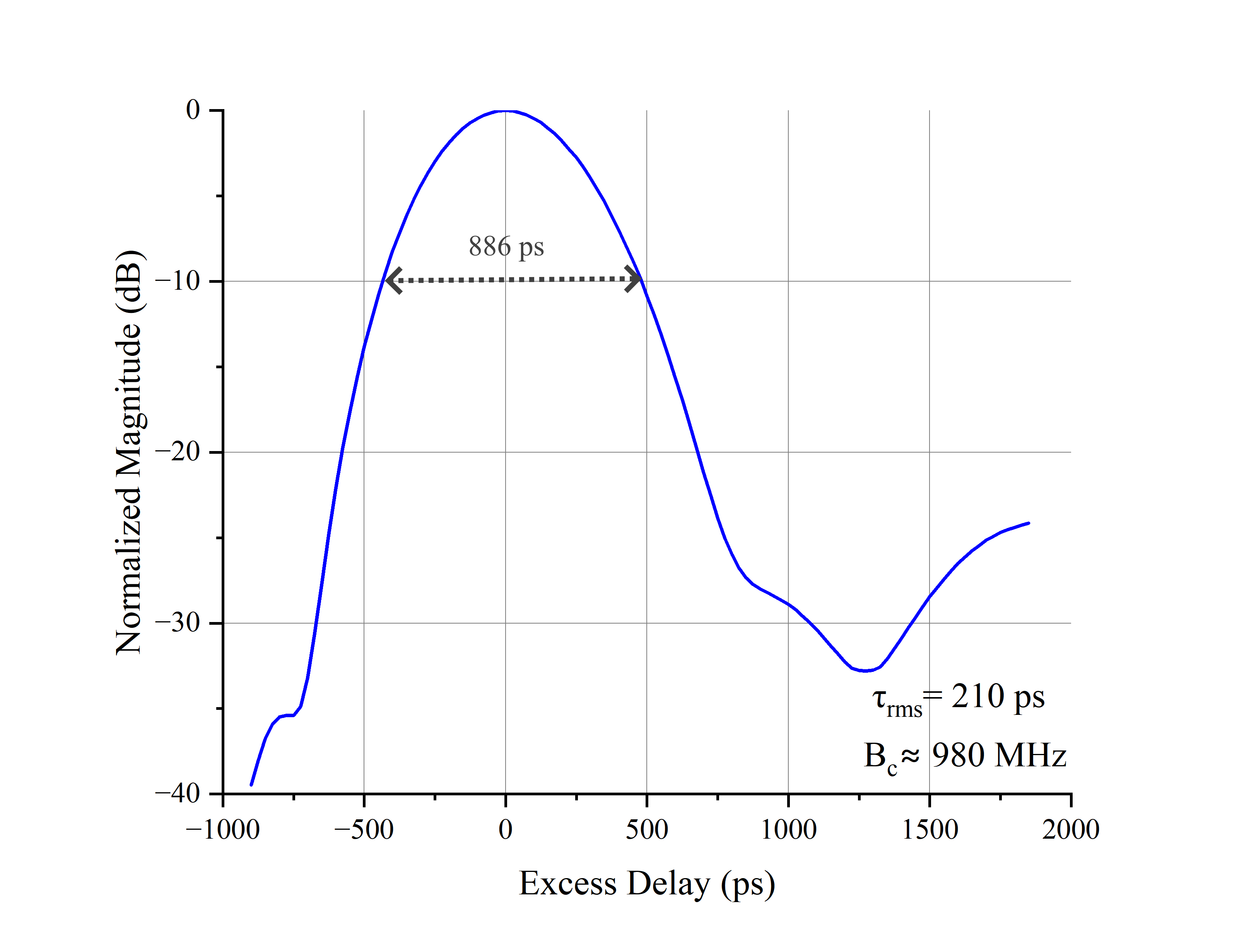}
\caption{Phantom-present channel impulse response obtained from measured
complex $S_{21}$. The measured system-response RMS delay spread is
approximately 210~ps.}
\label{fig:ir}
\end{figure}

For the sampled impulse response $h(\tau_i)$, the power-delay profile was
defined as
\begin{equation}
P(\tau_i)=\left|h(\tau_i)\right|^2.
\end{equation}
The mean delay $\bar{\tau}$ and root-mean-square (RMS) delay spread $\tau_{\mathrm{rms}}$
were calculated from the normalized moments
\begin{align}
\bar{\tau} & =
\frac{\sum_i P(\tau_i)\tau_i}{\sum_i P(\tau_i)}, \\
\tau_{\mathrm{rms}} & =
\sqrt{\frac{\sum_i P(\tau_i)(\tau_i-\bar{\tau})^2}
{\sum_i P(\tau_i)}}.
\end{align}
Multiplying $P(\tau_i)$ by a constant normalization factor does not alter
either moment because the factor cancels between the numerator and
denominator. The resulting measured system-response RMS delay spread is
approximately 210~ps. This value is of the same order as the nominal
250-ps resolution associated with the 4-GHz measurement bandwidth. 
The principal processing parameters and derived quantities are
summarized in Table~\ref{tab:channel}, and the resulting impulse response
is shown in Fig.~\ref{fig:ir}.
\begin{table}[!t]
\caption{Phantom-Present Time-Domain Processing and Results}
\label{tab:channel}
\centering
\footnotesize
\renewcommand{\arraystretch}{1.15}
\setlength{\tabcolsep}{4pt}

\begin{tabularx}{\columnwidth}{@{}p{0.40\columnwidth}X@{}}
\toprule
\textbf{Quantity} & \textbf{Value} \\
\midrule

Frequency span
& 4--8 GHz \\

Samples and spacing
& 201; 20 MHz \\

Window
& Hann \\

Transform
& Direct inverse fast Fourier transform (IFFT);
  \newline $8\times$ zero padding \\

Nominal Fourier resolution
& 250 ps \\

Delay interval
& 50 ns \\

Analysis gate and threshold
& 0--10 ns; $-20$ dB power \\

RMS delay spread
& Approximately 210 ps \\

\bottomrule
\end{tabularx}
\end{table}

\subsection{Absorption and Thermal Considerations}
\label{sec:safety}

\subsubsection{Numerical SAR Assessment}

The numerical dosimetry was obtained at 6.3~GHz from the final
two-probe CST full-wave model using the IEEE C95.3 1-g averaging
procedure. One probe was excited as the transmitter and the opposite
probe was matched as the receiver, consistent with the transmit--receive
configuration used for the $S_{21, dB}$ measurements. All SAR values were normalized to approximately 0.49 W accepted power at the two-port structure. The maximum 1-g-average SAR was $1.05\mathrm{W/kg}$.



\begin{figure}[htbp]
\centering
\includegraphics[width=0.5\textwidth]{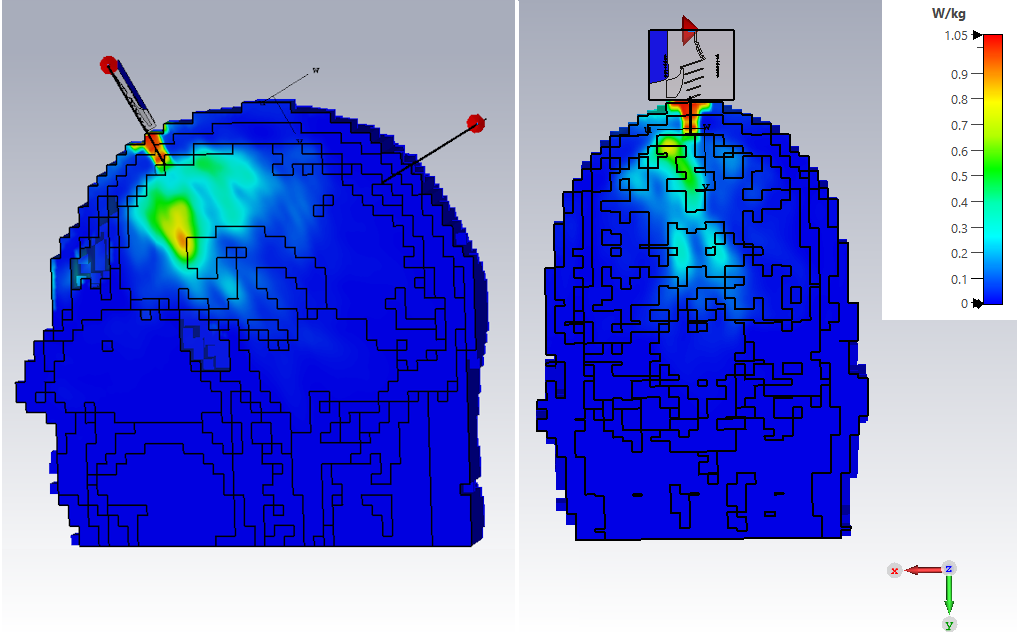}
\caption{Simulated 1-g-averaged SAR distribution for the two-probe
configuration at 6.3~GHz. For approximately 0.49 W accepted power, the maximum 1-g SAR was approximately 1.05 W/kg.}
\label{fig:sar_twoprobe}
\end{figure}

Fig.~\ref{fig:sar_twoprobe} shows the corresponding
simulated distribution. These are numerical results from the final two-probe CST model. 

The corresponding whole-tissue SAR was approximately 0.0164~W/kg. The
unrounded value, 0.0164164~W/kg, is internally
consistent with the simulated tissue power and mass,
$0.0920622~\mathrm{W}/5.60796~\mathrm{kg}=0.0164164~\mathrm{W/kg}$.

\subsubsection{Experimental Thermal Observation}

During phantom measurements, surface temperature in the coupling region
was monitored with a handheld infrared thermometer at an emissivity
setting of 0.95. Successive readings varied by approximately
$0.3^{\circ}\mathrm{C}$, and no systematic surface-temperature increase
was resolved.

\section{Discussion}
\label{sec:discussion}

\subsection{Principal Experimental Findings}

The controlled measurement establishes that introducing the finite,
brain-mimicking agar--gelatin--saline phantom into the retained
short-range two-probe
configuration changes the measured transmission response substantially.
Near 6.3~GHz, $|S_{21}|$ is approximately $-24$~dB with the
phantom present and approximately $-40$~dB with it absent, giving a
differential of approximately $+16$~dB. The corresponding
$|S_{21}|^2$ related ratio is approximately 39.8 for a 16-dB difference,
but this is not an absolute delivered-power ratio because the scattering
parameters are referenced to the VNA instrument ports.

The ex vivo goat-brain measurements provide a second experimental result.
Across the 5.78--6.86-GHz operating band, the goat-brain and phantom
$S_{21}$ responses closely follow one another, with an approximately
figure-derived maximum difference of about 1.5~dB. More than five
goat-brain repetitions produced reproducible general $S_{11}$ and
$S_{21}$ behavior. These measurements provide repeated ex vivo
validation of the operating-band response rather than formal
statistical analysis, without implying equivalence to human or in vivo
tissue.

\subsection{Physical Interpretation and Alternative Explanations}

The measured differential is a property of the complete finite
probe--medium electromagnetic system. Introducing the medium changes
loading, mutual coupling, interfaces, current distribution, propagation,
material loss, finite-boundary interactions, and the relationship between
the inserted wires and the external printed structures. Several of these
effects can occur simultaneously, and the present experiment does not
identify one as unique or dominant.

Consequently, the result must not be interpreted as intrinsic tissue gain,
material amplification, or evidence that bulk tissue attenuation is
smaller than established dielectric data imply.

The goat-brain agreement narrows, but does not eliminate, the range of
interpretations. 
It provides repeated ex vivo validation under nominally comparable measurement conditions. 

\subsection{Time-Domain Response and Bandwidth}

The time-domain response was obtained from the measured complex
phantom-present $S_{21}$ over 4--8~GHz using 201 points, 20-MHz spacing, a
Hann window, direct inverse Fourier transformation, and eightfold zero
padding. With a 0--10-ns analysis gate and a $-20$-dB power threshold, the
measured system-response RMS delay spread is approximately 210~ps. This
value is of the same order as the nominal 250-ps resolution associated
with the 4-GHz measurement bandwidth. 

\subsection{Relationship to Prior Work and Contribution}

Prior in-body channel studies have characterized path loss in homogeneous
tissue models and living-animal configurations, primarily at lower
frequencies~\cite{kurup2012,chavez2015,floor2015}. Warty \emph{et al.}
reported reflection and transmission measurements through a scalp
phantom for 2.45-GHz intracranial-pressure telemetry~\cite{warty2008}.
Brain-implantable antennas have also been studied in multilayer phantoms
over 2.45~GHz and 3--5~GHz~\cite{hout2019,sapari2022}. Kamel \emph{et
al.} measured implant-to-on-body transmission at 5.81~GHz with the
transmitting antenna immersed in chicken tissue~\cite{dualband2025}.

The principal result is not that the phantom behaves as a conventional attenuating transmission medium. Rather, introducing the finite medium changes the coupled electromagnetic boundary-value problem and substantially alters the measured two-port response.


\subsection{Practical Implications}

The measurements support evaluating candidate short-range biomedical
links as complete probe--medium--probe systems rather than screening carrier
frequencies from bulk attenuation alone. For this prototype, the useful
response is concentrated within the measured 5.78--6.86-GHz impedance
band and is reproduced qualitatively in ex vivo goat brain. Translation
requires mechanically controlled geometry, measured material properties,
probe-plane calibration or de-embedding, and validation in configurations
that more closely represent the intended deployment.

\section{Conclusion}
\label{sec:conclusion}

A printed tapered-slot coupling probe with a partially inserted 12-mm
wire element was characterized numerically and experimentally. Two
fabricated units exhibited a measured free-space $-10$dB band spanning 5.78
to 6.86~GHz, corresponding to a 1.08-GHz bandwidth and approximately
17.1$\%$ fractional bandwidth, with close agreement between the
measured and simulated reflection responses.

In the controlled two-probe experiment, the measured $|S_{21}|$ near
6.3~GHz was approximately $-24$~dB with the phantom present and
approximately $-40$~dB with it absent. The resulting approximately
$+16$~dB differential is interpreted as a medium-dependent
response of the complete probe--medium system, including loading, mutual
coupling, interfaces, geometry, propagation, and material loss. 
Numerical SAR for the two-probe transmit--receive model is reported in Section~\ref{sec:safety}.

More than five ex vivo goat-brain repetitions produced reproducible
general $S_{11}$ and $S_{21}$ behavior. Across 5.78--6.86~GHz, the
goat-brain $S_{21}$ response closely followed the phantom response, with
an approximately figure-derived maximum operating-band difference of
about 1.5~dB. This provides repeated ex vivo validation of the
operating-band response without establishing human, in vivo, or mechanistic
equivalence. 
Clinically, the probe confines the tissue-contacting element to a single thin wire, enabling minimally invasive placement, and agreement between the phantom and ex vivo data supports evaluating short-range intracranial links for wireless neural implants and intracranial sensors before animal studies.
Future work should use mechanically controlled geometry,
measured phantom dielectric properties, probe-plane calibration, and
additional biological configurations.



\section*{Conflict of Interest}

The authors declare no conflicts of interest.

\ifIEEEtranAvailable
  \bibliographystyle{IEEEtran}
\else
  \bibliographystyle{plain}
\fi
\bibliography{references}

\end{document}